\documentclass[twocolumn]{revtex4}

\usepackage{amsmath} 
\usepackage{amssymb} 
\usepackage{graphicx}
\usepackage{psfrag}

\newcommand{\id}{\mathbf{1}}

\newcommand{\suchthat}{\,|\,}
\newcommand{\set}[2]{ \{#1\suchthat #2\} }

\newcommand{\sset}[1]{ \{#1\} }
\newcommand{\vect}[1]{{\mathbf #1}}

\newcommand{\stab}{\mathcal S}
\newcommand{\cent}[1]{\mathcal Z(#1)}
\newcommand{\gauge}{\mathcal G}
\newcommand{\pauli}{\mathcal P}

\newcommand{\logical}{\mathcal L}

\newcommand{\rot}[1]{R_{#1}}

\newcommand{\qed}{\hspace*{\fill}$\blacksquare$}

\begin{document}

\title{Gauge Color Codes: \\Optimal Transversal Gates and Gauge Fixing in Topological Stabilizer Codes}

\author{H\'ector Bomb\'in}
\affiliation{Perimeter Institute for Theoretical Physics, 31 Caroline St. N., Waterloo, Ontario N2L 2Y5, Canada}
\affiliation{Deparment of Mathematical Sciences, University of Copenhagen, Universitetsparken 5, DK-2100 Copenhagen \O}

\begin{abstract}

Color codes are topological stabilizer codes with unusual transversality properties.
Here I show that their group of transversal gates is optimal and only depends on the spatial dimension, not the local geometry.
I also introduce a generalized, subsystem version of color codes.
In 3D they allow the transversal implementation of a universal set of gates by \emph{gauge fixing}, while error-dectecting measurements involve only 4 or 6 qubits.

\end{abstract}


\maketitle

\section{Introduction}

In fault-tolerant quantum computation~\cite{lidar:2013:quantum}, quantum information is protected from noise by encoding it in somewhat non-local degrees of freedom, thus distributing it among many smaller subsystems, typically qubits.
This makes sense under the physically relevant assumption that interactions with the environment have a local nature. 
The implementation of gates, consequently, must be also as local as possible to preserve the structure of noise.
This is naturally achieved with transversal gates, \emph{i.e.} unitary operators that transform encoded states by acting separately on suitable subsystems, in the simplest case independently on each qubit.
Unfortunately, no code admits a universal transversal set of gates, \emph{i.e.} such that transversal gates can approximate arbitrary gates~\cite{eastin:2009:restrictions}.


Topological quantum error correcting codes~\cite{kitaev:2003:ftanyons} introduce a richer notion of locality by considering the spatial location of the physical qubits, which are assumed to be arranged on a lattice.
They come in families parametrized with a lattice size, for a fixed spatial dimension.
Their defining features are (i) that the measurements needed to recover information about errors only involve a few neighbouring qubits and (ii) that no encoded information can be recovered without access to a number of physical qubits comparable to the system size.
Rather than sticking to transversal operations, for topological codes it is natural to consider instead local operations: quantum circuits of fixed depth with geometrically local gates~\cite{bombin:2012:universal,bravyi:2013:classification}.

Remarkably, spatial dimension constrains the gates that can be implemented by such local means. 
This is in particular true for topological \emph{stabilizer} codes, a popular class of codes for which it is often possible to obtain general results, as exemplified by the classification of 2D codes~\cite{bombin:2013:structure, yoshida:2011:classification, comment:higher}.
Along this line, it was shown recently~\cite{bravyi:2013:classification} that for $D$-dimensional topological stabilizer codes all local gates belong to the set $\pauli_D$, defined recursively~\cite{gottesman:1999:demonstrating} setting $\pauli_1$ to be the Pauli group $\pauli$ of operators and $\pauli_D$ to be the set of unitary gates $U$ with
\begin{equation}
U\pauli U^\dagger\subseteq\pauli_{D-1}.
\end{equation}
This result constraints the gates can be implemented locally, but unfortunately it says nothing about which valid gates can be realized in some topological stabilizer code.

Color codes are a class of topological stabilizer codes with remarkable transversality properties~\cite{kubica:2015:unfolding}.
The original color codes~\cite{bombin:2006:2dcc} were defined for $D=2$, with the aim of making the Clifford group $\pauli_2$ of gates transversal (see~\cite{nigg:2014:experimental} for a recent experimental implementation).
But color codes can also be defined for any $D>2$ on lattices called \emph{$D$-colexes}~\cite{bombin:2007:branyons}.
If colexes fulfilling certain local conditions can be constructed, families of color codes exist on dimension $D$ that admit the transversal implementation of the gates CNot and $\rot D:=\bigl(\begin{smallmatrix}
1&0\\ 0&e^{2\pi i/2^D}
\end{smallmatrix} \bigr)$~\cite{bombin:2007:3dcc,bombin:2013:self}.
Since $\rot D$ belongs to $\pauli_D$ but not to $\pauli_{D-1}$, color codes could saturate the geometrical constraint.

The difficulty is that it is not obvious that $D$-colexes with the required local conditions can be constructed for any $D$.
A first aim of this work is to show that actually such local conditions are irrelevant: the transversality properties of a color code are independent of the local geometry of the colex.
This is interesting at a theoretical level because $D$-colexes can be easily built for any dimension~\cite{bombin:2007:branyons} and thus the geometrical constraint on local gates is saturated \cite{comment:saturate}.
Notice in this regard that gates are not only local but also transversal.
On the practical side, with less constraints on the choice of the lattice of physical qubits more efficient codes are possible.
In 2D, for example, there exist families of color codes that for error correction only require measurements involving up to 6 physical qubits each.
With the constraints enforced, the number of qubits goes up to 8~\cite{bombin:2006:2dcc}.

Since no code admits a universal set of transversal gates, we are forced to find alternate routes.
A popular approach is the distillation of noisy \emph{magic states}~\cite{bravyi:2005:universal} (for which 2D color codes are well suited since the whole Clifford group is transversal).
It has the advantage of its quite general applicability, but the disadvantage that most resources end up being used for distillation, rather than the intended computation.
More efficient techniques are thus desirable, and in this regard two recent developments have been the \emph{gauge fixing} technique~\cite{paetznick:2013:universal} and the concatenation of different codes~\cite{jochym:2014:concatenated}.

But tricks to recover universal sets of gates have long been known.
In \cite{knill:1996:threshold} a code is considered such that the CNot and $R_3$ gates are transversal and the initialization and measurement in the computational and Hadamard rotated basis requires only transversal operations and error correction.
This suffices to complete the universal gate set with the Hadamard gate, at the price that each Hadamard requires an ancillary encoded qubit. 
The same technique applies to 3D color codes; indeed, it motivated their introduction.
From a practical perspective, however, 3D color codes pose two difficulties.
One is the mentioned requirement of ancillas.
The other is that the error-detecting measurements can involve each dozens of qubits: generally speaking, operations involving more qubits tend to be more unreliable and lengthy.

Both problems are removed in this work by introducing a subsystem form of color codes, \emph{gauge color codes}. 
Recall that in a conventional code quantum information is stored in a subspace of the Hilbert space corresponding to the physical qubits.
In a subsystem code, instead, this code subspace contains both logical and gauge qubits. 
The latter are just qubits that we do not care about, and in a topological code they might include local degrees of freedom.
This extra degrees of freedom can be put to work at least in two ways.
First, error detection measurements are potentially simplified by involving the gauge degrees of freedom~\cite{bombin:2010:subsystem}.
In the case of 3D gauge color codes, this materializes in measurements involving only 4 or 6 physical qubits, just as in 2D.
Second, as it was noticed recently~\cite{paetznick:2013:universal}, it might be possible to perform different transversal gates depending on the state of the gauge qubits.
This \emph{gauge fixing} technique applies to gauge color codes very neatly.
In particular for 3D gauge color codes it yields the same universal set of gates described above for conventional 3D color codes but without the need to use ancillary encoded states.
An important aspect is that gauge fixing is similar to error correction: it only requires local quantum operations supplemented with classical computation.

\section{Gates in stabilizer codes}

\subsection{Stabilizer codes}

Stabilizer codes~\cite{gottesman:1996:stabilizer,calderbank:1997:quantum} are a main object of study due to their balance of flexibility and simplicity.
Given a system of $n$ qubits, a \emph{stabilizer} subgroup $\stab\subseteq\pauli$, with $-\id\not\in\stab$, defines a subspace, or code, of states $\psi$ with $s\psi =\psi$ for every $s\in\stab$.
A \emph{subsystem} stabilizer code~\cite{poulin:2005:stabilizer} is defined by giving in addition a \emph{gauge} group $\gauge\subseteq\pauli$ such that $\stab$ is the center of $\gauge$ up to phases.
The subspace stabilized by $\stab$ splits in two subsystems: the gauge group generates the full algebra of operators on one of them and acts trivially on the other.
Logical qubits (those to be protected) inhabit the later, gauge qubits the former.
The elements of $\cent \gauge$, \emph{i.e.} the group of Pauli operators that commute with the elements of $\gauge$, are called \emph{bare} logical operators: they only act on logical qubits.
Elements of $\cent\stab$ are their \emph{dressed} counterpart: they may act on gauge qubits.
Both quotients $\cent\gauge/\stab$ and $\cent\stab/\gauge$ yield the Pauli group on logical qubits.

\subsection{Transversal gates}
 
Let $X_q$, $Z_q$ denote the Pauli operators $X$, $Z$ acting on the qubit  $q$, and similarly
\begin{equation}
X_S:=\prod_{q\in S}X_q,\qquad Z_S:=\prod_{q\in S}Z_q,
\end{equation}
for a set $S$ of qubits.
Of interest here are stabilizer codes with (i) a CSS structure~\cite{calderbank:1996:good,steane:1996:multiple}: $\gauge$ has a generating set $\gauge_0$ such that each of its elements takes the form $X_S$ or $Z_S$ for some $S$, and (ii) a single encoded qubit with bare logical operators $X_Q$, $Z_Q$, where $Q$ is the set of all physical qubits.
For such codes the CNot gate is trivially transversal, and also the Hadamard gate if in addition the code is self-dual, \emph{i.e.} $X_S\in\gauge$ if and only if $Z_S\in\gauge$.

More interesting is the gate $\rot n=
\bigl(\begin{smallmatrix}
1&0\\ 0&e^{2\pi i/2^n}
\end{smallmatrix} \bigr)$.
Let $|S|$ denote the cardinality of a set $S$.
As shown in appendix~\ref{app:transversal}, $\rot n$ is transversal if the set $Q$ can be divided into two disjoint sets $T$ and $T'$, \emph{i.e.} $Q=T\sqcup T'$, such that for every collection of generators $X_{S_1},\dots,X_{S_m}\in \gauge_0$, where $1\leq m\leq n$,
\begin{equation}\label{eq:condition}
\left|T\cap\bigcap_{i=1}^m S_i\right |\equiv \left|T'\cap\bigcap_{i=1}^m S_i\right |\mod 2^{n-m+1}.
\end{equation}

\subsection{Gauge fixing}

The idea of the gauge fixing technique~\cite{paetznick:2013:universal} is that by fixing some of the gauge degrees of freedom to certain values it might be possible to recover transversal gates that would otherwise be forbidden.
A code $\stab_2,\gauge_2$ is a gauge fixed version of  $\stab_1,\gauge_1$ if $\stab_2$ extends $\stab_1$ by fixing the values of some operators in $\gauge_1$, \emph{i.e.}
\begin{equation}\label{eq:fix}
\stab_1\subseteq \stab_2\subseteq\gauge_1,\qquad \gauge_2\propto\cent{\stab_2}\cap \gauge_1.
\end{equation}
The relations \eqref{eq:fix} become most transparent by choosing canonical generators $X_i, Z_i$ of the Pauli group, $i=1,\dots,n$ such that
\begin{align}
\stab_1&=\langle Z_1,\dots, Z_{r}\rangle,\\
\stab_2&=\langle Z_1,\dots,Z_{s}\rangle,\\
\gauge_1&\propto\langle Z_1,\dots,Z_r,X_{r+1},Z_{r+1}\dots,X_t,Z_t\rangle,
\end{align}
where $r\leq s\leq t\leq n$.
Computing centralizers is a trivial task, \emph{e.g.}
\begin{equation}
\cent{\gauge_1}\propto\langle Z_1,\dots,Z_r,X_{t+1},Z_{t+1}\dots,X_n,Z_n\rangle.
\end{equation}
In particular, as canonical generators of the logical Pauli group for the code $\stab_1,\gauge_1$ one can take the following representatives of $\cent{\gauge_1}/\stab_1$
\begin{equation}\label{eq:logical}
X_{t+1},Z_{t+1},\dots,X_{n},Z_{n}.
\end{equation}
Now, according to \eqref{eq:fix} we should take
\begin{equation}
\gauge_2\propto\langle Z_1,\dots,Z_s,X_{s+1},Z_{s+1}\dots,X_t,Z_t\rangle.
\end{equation}
This is indeed a valid choice: $\stab_2$ is the center of $\gauge_2$ up to phases.
Moreover, the operators \ref{eq:logical} are clearly also canonical generators of the logical Pauli group for the code $\stab_2,\gauge_2$.


Since $\stab_1\subseteq\stab_2$ and the two codes share logical operators, encoded states of the second code can be regarded as being encoded states of the first.
Conversely, given an encoded state of the first code it is possible to make it into an encoded state of the second code without affecting the logical operators.
This amounts to fix the eigenvalues of $Z_{r+1},\dots,Z_{s}$ to +1, which can be done in two steps.
First the $Z_{r+1},\dots,Z_{s}$ need to be measured, or equivalently any other generators of $\stab_2/\stab_1$.
Then a product of a subset of the operators $X_{r+1},\dots,X_{s}$ is applied, or equivalently any other generators of $\gauge_1/\gauge_2$.
In particular, $X_i$ is applied if the measurement yields $Z_i=-1$.
All the operations involved commute with the logical operators \eqref{eq:logical}, and therefore are safe to perform.

An alternative characterization of~\eqref{eq:fix} is possible in terms of logical operators.
Suppose that the two codes share a set $\logical$ of representatives of bare logical operators.
Then \eqref{eq:fix} holds (up to irrelevant phases in $\gauge_1$) if and only if
\begin{equation}\label{eq:fix1}
\stab_1\subseteq\stab_2.
\end{equation}
Alternatively, under the same assumption \eqref{eq:fix1} holds (up to a choice of signs in the stabilizers and irrelevant phases in $\gauge_1$) if and only if
\begin{equation}\label{eq:fix2}
\gauge_2\subseteq\gauge_1.
\end{equation}
To check the statements (\ref{eq:fix1},\ref{eq:fix2}), notice first that for any code $\stab,\gauge$ a set $\logical$ of representatives of bare logical operators satisfies 
\begin{equation}\label{eq:cent}
\stab\propto\cent{\logical\gauge}, \qquad \gauge\propto\cent{\logical\stab}.
\end{equation}
The equivalence of~\eqref{eq:fix1} and~\eqref{eq:fix2} under the assumption that the two codes share such a set $\logical$ follows from the basic properties~\eqref{eq:cent}. 
Indeed, \eqref{eq:fix2} implies that
\begin{equation}
\stab_1\propto\cent{\logical\gauge_1}\subseteq\cent{\logical\gauge_2}\propto \stab_2,
\end{equation}
and conversely \eqref{eq:fix1} implies
\begin{equation}
\gauge_2\propto\cent{\logical\stab_2}\subseteq\cent{\logical\stab_1}\propto \gauge_1,
\end{equation}
According to the above discussion, if~\eqref{eq:fix} is satisfied there exist such a shared set $\logical$ and \eqref{eq:fix1} holds.
Conversely, if $\logical$ is shared and~\eqref{eq:fix1} holds then up to phases
\begin{equation}
\stab_2\propto\gauge_2\cap\cent{\gauge_2}\subseteq\gauge_2\subseteq\gauge_1,
\end{equation}
which used~\eqref{eq:fix2} and completes the first part of~\eqref{eq:fix}, and
\begin{multline}
\cent{\stab_2}\cap\gauge_1\propto\cent{\stab_2}\cap\cent{\logical\stab_1}=\\=\cent{\logical\stab_1\stab_2}=\cent{\logical\stab_2}\propto\gauge_2,
\end{multline}
which is the second part of~\eqref{eq:fix} .

\section{Color codes}

\subsection{Simplicial lattice}
 
 \begin{figure}
  \centering
  \includegraphics[width=7.5cm]{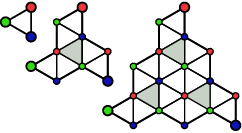}
  \caption{
    The first three instances of an infinite family lattices for 2D color codes.
  A qubit is attached to each triangle, thick link and large vertex.
  Triangles in the set $T$ are shadowed.
  }
\label{fig:2DCC}
\end{figure}

Color codes require lattices called colexes~\cite{bombin:2007:branyons}, but here the focus is shifted to their dual lattices. 
For those readers already familiar with colexes, the correspondence is clarified in appendix~\ref{app:duality}.

For simplicity we focus on the 2D case.
The key to the construction are triangles (2-simplices) with the vertices (0-simplices) labeled with 3 given colors (thus each side or 1-simplex is labeled by 2 colors).
We are interested in simplicial lattices (triangulations) with the overall shape of such a colored triangle $M$.
The lattices must have 3-colored 0-simplices such that (i) each 2-simplex is properly colored and (ii) the sides of $M$ have the same coloring as the 1-simplices that lie on them. 
In Fig.~\ref{fig:2DCC} the first examples of an infinite sequence of such lattices is given: they are obtained by cutting triangles of different sizes out of the same infinite triangular lattice.

\begin{figure}
  \centering
  \includegraphics[width=3cm]{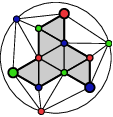}
  \caption{
Triangulation of a sphere built from a `triangle' $M$ (shadowed): the hidden half of the sphere is a 2-simplex.
}
\label{fig:close}
\end{figure}

From every such triangulation one can get a 3-colored triangulation of a sphere just by adding 3 extra 0-simplices as shown in Fig.~\ref{fig:close}: there is one new 2-simplex for each 1-simplex on the boundary of $M$, one for each of the vertices of $M$, and one formed by the 3 new 0-simplices.
It is out of this triangulation of the sphere that a color code can be built, in particular via certain sets $\Delta_d$: $\Delta_d$ contains all d-simplices in the new triangulation except those that do not contain any of the original 0-simplices of $M$.
Notice in particular that the elements of $\Delta_2$ correspond one to one to the following elements of $M$: all 2-simplices, 1-simplices lying on the sides of M, and 0-simplices that are also vertices of $M$.

The general $D$-dimensional case is entirely analogous, with $D$-simplices instead of triangles and $D+1$ colors instead of 3.
In particular, every $d$-simplex is labeled by $d+1$ colors.
For a formal exposition see appendix~\ref{app:duality}.

\begin{figure}
  \centering
  \includegraphics[width=8cm]{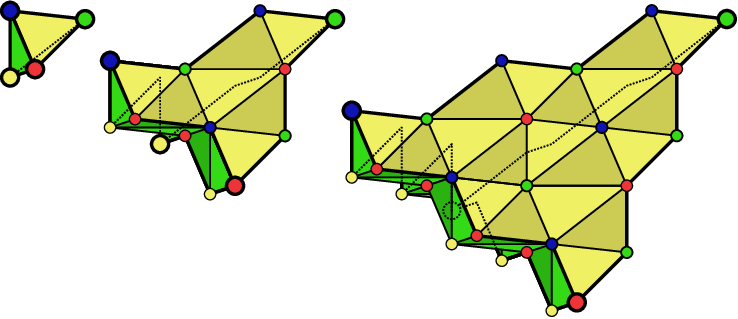}
  \caption{
  The first three instances of an infinite family of 3D color codes.
  Qubits are attached to all 3-simplices, to 2-simplices covering the external faces, to 1-simplices displayed thicker and to 0-simplices displayed larger.
 2-simplices are colored with the complementary to their 3 colors.
 Those that are darker belong to the set $T_2$.
  }
\label{fig:3DCC}
\end{figure}

The 3D case is the most important, so here is a construction where a tetrahedron is carved out of a suitably colored BCC lattice.
Vertices are placed at points $\vect x$ or $\vect x+\vect l_0$, where $\vect x$ has integer coordinates and $\vect l_0:=(\frac 1 2,\frac 1 2,\frac 1 2)$.
Red, green, blue and yellow vertices have positions $\vect x$ such that $\vect x\cdot\vect l_0$ is $0$, $\frac 1 2$, $\frac 3 4$ and $\frac 1 4$ modulo 1, respectively.
Any 3-simplex has vertices of the form 
\begin{equation}\label{eq:vertices}
\vect x,\,\,\, \vect x+\vect a, \,\,\,\vect x+\frac 1 2( \vect a+s\vect b+\vect c), \,\,\,\vect x+\frac 1 2( \vect a+s\vect b-\vect c),
\end{equation}
where $s=\pm 1$ and the triad $(\vect a,\vect b,\vect c)$ is a permutation of the triad $(\vect i,\vect j,\vect k)$ of canonical vectors.
Given a positive integer $n$, retain those simplices with vertices $\vect x$ such that
\begin{equation}
\vect x\cdot \vect l_k\leq \frac k 4+(n-1)\,\delta_{k0}, \qquad k=0,1,2,3,
\end{equation}
where  $\vect l_1:=(\frac 1 2,-\frac 1 2,-\frac 1 2)$, $\vect l_2:=(-\frac 1 2,\frac 1 2,-\frac 1 2)$, $\vect l_3:=(-\frac 1 2,-\frac 1 2,\frac 1 2)$.
This gives a tetrahedron: each constraint produces a face with coloring dictated by $k$, see Fig.~\ref{fig:3DCC}.

\subsection{Gauge color codes}
 
Consider sets of $d$-simplices $\Delta_d$ from a $D$-dimensional construction as above, $D\geq 2$.
Attach a physical qubit to each $D$-simplex in $\Delta_D$, also denoted $Q$, and define the groups of operators
\begin{equation}\label{eq:groupsC}
C(d,e):=\langle \set{X_{S_\delta}}{\delta\in\Delta_{d-1}}\cup \set{Z_{S_\delta}}{\delta\in\Delta_{e-1}}\rangle,
\end{equation}
where $e,d$ are positive integers, $d,e<D$, $S_\delta$ is the set of $D$-simplices that contain $\delta$, and $\langle A \rangle$ denotes a group with generators $A$.
The groups \eqref{eq:groupsC} satisfy
\begin{equation}\label{eq:partial}
C(d_1,e_1)\subseteq C(d_2,e_2)\iff d_1\leq d_2, e_1\leq e_2,
\end{equation}
and their centralizer takes the form
\begin{equation}
\cent {C(d,e)}\propto \langle X_Q, Z_Q \rangle \cdot C(\bar e,\bar d), \label{eq:ZC}
\end{equation}
where $\propto$ denotes equality up to global phase~\cite{bombin:2013:self}.
Moreover, $|Q|$ is odd and the group $\langle X_Q, Z_Q \rangle$ has trivial intersection with any of the groups $C(d,e)$.

For positive integers $d, e$ with $d+e\leq D$, the $(d,e)$ gauge color code is defined by
\begin{equation}
\stab:=C(d,e),\quad \gauge:=C(\bar e,\bar d) \label{gauge_gen}
\end{equation}
where $\bar x:=D-x$.
According to the above properties the definition is valid: $\stab$ is abelian and the center of $\gauge$.
Whatever the values of $d,e$, there is a single encoded qubit with logical Pauli operators $X_{Q}, Z_{Q}$.
When $e=\bar d$ the color code is conventional~\cite{bombin:2013:self}.
Conventional color codes yield all the possible quotient groups appearing in the construction of logical operators.
In particular, it follows from the results on conventional color codes of \cite{bombin:2007:branyons} that logical operators are always non-local: those of the form $X_S$ are $\bar d$-brane-net like if bare and $e$-brane-net like if dressed, and those of the form $Z_S$ are $\bar e$-brane-net like if bare and $d$-brane-net like if dressed.

Some properties of the above 2D and 3D examples can be extracted from the first instances of each family.
For example, in the 2D case the number of qubits is $1+3 n+3 n^2$, and in the 3D case it is $1+4 n+6 n^2+4 n^3$.
Also, in 2D the generators of $C(1,1)$ have support on at most 6 qubits, and the same holds for $C(2,2)$ in 3D.

\subsection{Transversality}

For gauge color codes the CNot gate is transversal.
For self-dual codes, $d=e$, the Hadamard gate is transversal.
What about the gate $\rot n$?
As shown in appendix~\ref{app:transversal_cc}, a set $T$ as in \eqref{eq:condition} exists if
\begin{equation}\label{eq:condition_D}
D\geq n\bar e.
\end{equation}
In particular, $\rot D$ can be implemented by taking $\bar e=1$.
The result \eqref{eq:condition_D} was obtained for $T=\emptyset$ and $d=\bar e$ in \cite{bombin:2013:self}, but an important limitation there was the lack of a general recipe to construct lattices satisfying \eqref{eq:condition}.

As it turns out the set $T$ only depends on the lattice, not the specific $(d,e)$ parameters of the code. 
It must be such that the operator $X_T$, regarded as an error in the corresponding $(1,D-1)$ color code, commutes with a $Z$ stabilizer generator if and only if the generator has support on a number of qubits that is a multiple of 4.

For the above 2D and 3D families of color codes we can give $T$ explicitly.
In the 2D case it is simplest to depict it, see Fig.~\ref{fig:2DCC}.
As for the 3D case, first notice that qubits correspond to the following elements of the triangulation of a colored tetrahedron $M$: all 3-simplices, the 2-simplices on the faces of $M$, the $1$-simplices on the edges of $M$, and the $0$-simplices that are the vertices of $M$.
A valid choice for the set $T$ is the union of (i) the set $T_3$ of all 3-simplices with $(\vect a,\vect b,\vect c)$ as in \eqref{eq:vertices} an even permutation of the triad $(\vect i,\vect j,\vect k)$, (ii) the set $T_2$ of all 2-simplices that are not a subsimplex of any element of $T_3$, see Fig.~\ref{fig:3DCC}, and (iii) the set $T_1$ containing all 1-simplices that are not a subsimplex of any element of $T_2$.

\subsection{Gauge fixing}

The different gauge color codes that can be defined on a given, fixed lattice are related by gauge fixing, either directly or via some other code in the family.
Indeed, all the codes have a shared group $\logical$ of bare logical operators (namely, those with support in all qubits) and the condition \eqref{eq:fix2} is satisfied when $d_1\leq d_2$, $e_1\leq e_2$.
Take \emph{e.g.} $D=3$.
For a given geometry we might consider both the code $(1,1)$ or the gauge fixed version $(1,2)$.
Within $(1,1)$ CNot and Hadamard gates are transversal.
Fixing the gauge we can move into $(1,2)$ to apply transversal $\rot 3$ gates, completing a universal set of transversal gates.
The ideal strategy is to transition to $(1,2)$ only momentarily.
This way there is no need to ever measure directly the large stabilizer generators of the $1,2)$ code.
Instead, it is enough to measure the gauge generators in $C(2,2)$, which only involve up to 6 qubits.
Notice that the only non-transversal element of the procedure is the classical computation to find the gauge operator that will fix the gauge as desired.
This is entirely analogous to error correction.

\section{Measurements in error correction}\label{app:correction}

In order to perform error correction on a stabilizer code the first step is to recover the error syndrome by measuring a collection of stabilizer operators that generate $\stab$.
This can be done either directly or indirectly by performing a sequence of measurements of gauge operators from which the syndrome can be recovered~\cite{bombin:2010:subsystem}.
In the case of CSS codes the later approach is particularly straightforward.
It suffices to first measure a generating set of $X$-type gauge operators (which commute with each other), and then do the same with $Z$-type operators.
The eigenvalues of $X$-type stabilizers can be recovered from the first set of measurements, and similarly for the $Z$-type and the second set.

Consider now the specific case of a $(d,e)$ gauge color code.
From the $D$-colex perspective in order to measure the syndrome for $Z$-type stabilizers we have to either measure directly stabilizer operators attached to $(\bar e+1)$-cells or measure instead gauge generators attached to $(d+1)$-cells.
In fact, it is possible to choose to measure operators attached to $d'$-cells for any $d+1\leq d'\leq \bar e+1$).
Whenever $d'<\bar e+1$, it is worth noting that it is not necessary to measure \emph{all} the $d'$-cell operators.
Instead, the stabilizer generator $Z_c$ on a $(\bar e+1)$-cell $c$ can be recovered in different ways from the operators $Z_{c'}$, where $c'$ stands for a $d'$-cell.
In particular, if $\kappa$ is a subset with $d'$ elements from the set of $\bar e+1$ colors of $c$, and $c_\kappa$ the set of $\kappa$-cells contained in $c$, then~\cite{bombin:2007:branyons}
\begin{equation}\label{eq:decompose_cell}
Z_c=\prod_{c'\in c_\kappa}Z_{c'}
\end{equation}
which is just a general form of \eqref{eq:constraint_local}.

It follows that it is enough to measure at a subset of $d'$-cells with enough color combinations $\kappa$ so that every collection of $\bar e+1$ colors has to contain one of the color combinations.
\emph{E.g.} for $(1,1)$ gauge color codes in 3D it suffices to choose two disjoint pair of colors, as opposed to the six possible pairs of colors.

Consider more particularly the specific 3D lattice given in the main text.
In its bulk, the 3-cells of the corresponding 3-colex have 24 qubits (0-cells) each.
For a specific pair of colors, each cell has either 6 2-cells with 4 qubits each or 4 2-cells with 6 qubits each.
Measuring any of the corresponding sets of 2-cell operators is enough, but measuring all of them gives redundant information that can be put to good use~\cite{bombin:2014:single-shot}.

Coloring can be used, whatever the spatial dimension $D$, to organize the measurements.
The idea is that cells with the same coloring have disjoint sets of vertices, and thus the related operators act on disjoint set of qubits.
Then one can measure in parallel all the generators related to a color combination.
Naturally, other approaches might be more optimal.

Finally, a note on error-correction itself.
In CSS codes it is natural (but again not optimal) to deal separately with $X$ and $Z$ errors.
In the case of gauge color codes this has the advantage of mapping the problem back to conventional color codes.
Namely, consider a $(1,1)$ gauge color code in 3D.
The error syndrome for $X$ errors is the same as in the corresponding $(2,1)$ conventional color code, and the error syndrome for $Z$ errors is the same as in the corresponding $(1,2)$ conventional color code.
In the presence of measurements error, however, new scenarios open if gauge generators are measured~\cite{bombin:2014:single-shot}.

\section{Outlook}

The results presented here show that 3D gauge color codes put together some unique features (in fact, they turn out to be surprisingly resilient to errors in the error-detecting measurements, as explained in~\cite{bombin:2014:single-shot}).
Further research would thus be desirable, \emph{e.g.} regarding noise thresholds.

It is intriguing to consider quantum Hamiltonian models based on 3D gauge color codes, \emph{i.e.} of the form \begin{equation}
H=\sum_{g\in 
\gauge_0} J_g \, g
\end{equation}
for some set $\gauge_0$ of local generators of the gauge group and couplings $J_g$.
The fact that all the (standard) gauge generators detect fluxes~\cite{bombin:2007:branyons} suggests the possibility of a self-correcting phase~\cite{dennis:2002:tqm,brown:2014:qm}.

Topological codes have a rich behavior, but they are just part of the larger class of local codes, where spatial geometry is not relevant.
This could be a path to obtain code families with the properties of 3D gauge color codes but requiring less physical qubits.



I am grateful to Stephen Bartlett, Steve Flammia and Courtney Brell for pointing out the work~\cite{paetznick:2013:universal} in connection with color codes.
I thank the Spanish MIC-INN grant FIS2009-10061, CAM QUITEMAD S2009-ESP-1594, the Sapere Aude grant of the Danish Council for Independent Research, the ERC Starting Grant QMULT and the CHIST-ERA project CQC.
Work at Perimeter Institute is supported by Industry Canada and Ontario MRI.


\appendix

\section{Transversal gates}\label{app:transversal}

The aim is to show that $\rot n$ is transversal if \eqref{eq:condition} holds, which can be written more compactly as
\begin{equation}\label{eq:assumption}
\left|\bigcap_{i=1}^m S_i\right|_T\equiv 0 \mod 2^{n-m+1}
\end{equation}
by introducing the notation, for any subset $A\subseteq Q$
\begin{equation}
|A|_T:=|T'\cap A|-|T\cap A|.
\end{equation}

\noindent In particular, the claim is that there exists an integer $k$ such that a logical $\rot n$ is implemented by applying $\rot n^{-k}$ to qubits in $T$ and $\rot n^k$ to the rest, \emph{i.e.} in $T'$. 
The resulting gate maps, for any given subset $A\subseteq Q$,
\begin{equation}\label{eq:transform}
X_{A}|\vect 0\rangle \rightarrow \exp\left (\frac{2k\pi i}{2^n}|A|_T\right) X_{A}|\vect 0\rangle
\end{equation}
where $|\vect 0\rangle$ is the state with all physical spins up $|0\rangle^{\otimes |Q|}$.
In the code subspace the stabilizers have eigenvalue 1 and thus encoded states are superpositions of states $X_{A}|\vect 0\rangle$ with $X_{A}$ a product of several $X$-type gauge generators and possible $X_Q$.
Taking $Z_Q$ to be the logical $Z$ operator, an encoded state $|a\rangle$, $a=0,1$, is a superposition of states of the form
\begin{equation}\label{eq:states}
X_Q^aX_G|\vect 0\rangle =
\begin{cases}
   X_G|\vect 0\rangle & \text{if } a = 0 \\
   X_{Q- G}|\vect 0\rangle & \text{if } a=1
 \end{cases}
\end{equation}
where $X_G\in\gauge$ and $Q-G$ is the complement of $G$ in $Q$.

Suppose that, for any $X_G\in\gauge$, we had
\begin{equation}\label{eq:goal}
|G|_T\equiv 0\mod 2^{n}.
\end{equation} 
Then the states \eqref{eq:states} transform as follows according to \eqref{eq:transform}
\begin{align}
   X_G|\vect 0\rangle & \rightarrow X_G|\vect 0\rangle\\
   X_{Q-G}|\vect 0\rangle & \rightarrow \exp\left(\frac {2k\pi i|Q|_T}{2^n}\right) X_{Q-G}|\vect 0\rangle,
\end{align}
using the fact that for sets $A,B,C$ with $A\subseteq B$
\begin{multline}
|C\cap (B-A)|=|(C\cap B)-(C\cap A)|=\\=|C\cap B|-|C\cap A|.
\end{multline}
Since $X_Q$ and $Z_Q$ anticommute $|Q|$ is odd and so is $|Q|_T$ too.
Therefore $|Q|_T$ and $2^n$ are relatively prime and there exists $k$ with 
\begin{equation}
k|Q|_T\equiv 1\mod 2^n.
\end{equation}
$R_n$ is indeed implemented for such $k$.

Thus it suffices to show that \eqref{eq:goal} holds.
For $G=\emptyset$ this is trivial.
For $G\neq\emptyset$ consider the stronger statement
\begin{equation}\label{eq:toproof}
\left|G\cap\bigcap_{i=1}^m S_i\right |_T\equiv 0\mod 2^{n-m}
\end{equation}
for any $m=0,\dots n$, $X_{S_i}\in\gauge_0$, which reduces to \eqref{eq:goal} for $m=0$.
Let 
\begin{equation}
X_G=\prod_{i=1}^r X_{G_i}
\end{equation} for some $X_{G_i}\in\gauge_0$.
For $r=1$ \eqref{eq:toproof} is true by assumption \eqref{eq:assumption}. 
If $r>1$, set 
\begin{equation}
X_{G'}=\prod_{i=1}^{r-1}X_{G_i}.
\end{equation}
Then $G=G'+S_r$, with $+$ the symmetric difference, defined as
\begin{equation}
A+B:=(A\cup B)-(A\cap B).
\end{equation}
Noticing that
\begin{equation}
|A\cap (B+C)|=|A\cap B|+|A\cap C|-2|A\cap B\cap C|
\end{equation}
one immediately gets
\begin{multline}
\left|G\cap\bigcap_{i=1}^m S_i\right |_T=\left|G'\cap\bigcap_{i=1}^m S_i\right |_T+\left|G_r\cap\bigcap_{i=1}^m S_i\right |_T+\\-2\left|G'\cap \left (G_r \cap \bigcap_{i=1}^m S_i\right )\right |_T.
\end{multline}
The result follows by induction on the number of generators $r$: all the terms on the right hand side involve less than $r$ generators, in particular $r-1$ in the case of $G'$ and $1$ in the case of $G_r$, and
\begin{equation}
x\equiv 0\mod 2^{n-(m+1)}\Rightarrow 2x\equiv 0\mod 2^{n-m}
\end{equation}

A similar result holds for $k$ control qubits with $m$ running from $k+1$ to $k+n$ and the modulus being $2^{n-m+k+1}$.

\section{$D$-colexes and duality}\label{app:duality}

$D$-colexes are $D$-dimensional lattices with certain colorability properties.
For a lattice here it is meant a division of a closed $D$-manifold into $D$-cells, which are themselves composed of 0-cells (vertices), 1-cells (edges) and so on in the usual way.
The colorability reads: 

\vspace{.2cm}
\noindent\emph{Every $d$-cell is labeled by $d$ colors from a given set of $D+1$ colors. Given a cell $c$ with color set $\kappa$, the cells with $c$ as a subcell are in one to one correspondence (according to their label) with the color sets with $\kappa$ as a subset.}
\vspace{.2cm}

\noindent\emph{E.g.} at every vertex $D+1$ edges meet, each with a different color.
Notice that every two cells with the same color set must be equal or disjoint.
This in turn implies that all 2-cells have an even number of edges: the edges along the boundary of the 2-cell must have alternate colors.
Another easy property is that the boundary of every $d$-cell is itself a $(d-1)$-colex (in particular a $(d-1)$-sphere).

The dual of a $D$-colex is a simplicial lattice with the vertex colorability properties given in the text, in particular, the dual of a $d$-cell with color set $\kappa$ is labeled with the color set $\bar \kappa$, defined as the complement of $\kappa$ in the set of $D+1$ colors.
Recall that under duality $d$-cells are mapped to $\bar d$-cells in such a way that the relationship `is a subcell of' is reverted.
It is then easy to check that one indeed gets a simplicial lattice with the right coloring: each of the subsimplices of a given simplex with color set $\kappa$ is labeled by a nonempty subset of $\kappa$.

The above applies to closed $D$-manifolds, but of interest here are punctured $D$-colexes.
They are obtained by starting from a $D$-colex on a $D$-sphere and removing a single vertex, together with all the cells that contain it.
In the dual simplicial lattice this means removing a $D$-simplex together with all of its subsimplices, giving rise to the simplex collections $\Delta_d$ of the main text.
But some of the simplices in $\Delta_d$ have subsimplices that are not in any of the $\Delta_d$.
The simplicial lattice $M$ of the main text is recovered by keeping instead only those simplices that still retain all their subsimplices.

\begin{figure}
  \centering
  \includegraphics[width=8.3cm]{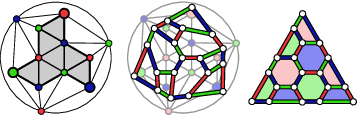}
  \caption{
A punctured 2-colex with faces colored with their complementary/dual color (right) compared to the corresponding triangulation of a sphere (left). Duality is explicit in the central figure.
}
\label{fig:duality}
\end{figure}

All this is illustrated for $D=2$ in Fig.~\ref{fig:duality}, where $M$ (shaded), its closed spherical version and the corresponding punctured colex are compared.
It is apparent that the colex picture allows a more easy visualization of the code, with qubits placed at vertices.
Notice that in the $D$-colex perspective the generators of stabilizer and gauge group correspond to the following geometrical objects: for a $(d,e)$ gauge color code $X$ stabilizer generators are $(\bar d+1)$-cells, $Z$ stabilizer generators are $(\bar e+1)$-cells, $X$ gauge generators are $(e+1)$-cells and $Z$ gauge generators are $(d+1)$-cells.

Another interesting point is that when we describe color codes by giving $M$ it is obvious what the simplest example is: that in which the triangulation of $M$ is composed of just a single $D$-simplex that coincides with $M$ itself.
From the colex perspective this corresponds to a punctured $D$-colex that, without the vertex removed, is the boundary of a $(D+1)$-cube. 
It suffices to color parallel edges of the cube with the same color to get the colex structure, which is thus a $D$-sphere as required.
The dual simplicial picture of this simplest case serves as a pattern for the combinatorial prescription given in the main text for `closing' a generic triangulation of $M$ by adding extra simplices: it shows that the prescription indeed produces a $D$-sphere as stated.

\section{Perfect colexes}\label{app:perfect}

\subsection{Motivation}\label{app:perfect_motivation}

The result~\eqref{eq:condition_D} was already proven in~\cite{bombin:2013:self} for conventional color codes in the special case $T=\emptyset$, \emph{i.e.} when the same rotation is applied to all physical qubits.
Generalizing slightly to the subsystem case, this means that the result hinged on gauge generators $X_{S_i}$ satisfying
\begin{equation}\label{eq:condition_empty}
\left|\bigcap_{i=1}^m S_i\right|\equiv 0 \mod 2^{n-m+1}.
\end{equation}
These are local constraints on the $D$-colex, because the generators $X_{S_i}$ are local.

Fortunately conditions \eqref{eq:condition_empty} can be simplified a lot.
This is due to the following property of colexes~\cite{bombin:2013:self}:

\vspace{.2cm}
\noindent \emph{Given a collection of cells $c_i$ with color sets $\kappa_i$, their intersection is a collection of cells with color label $\kappa$:}
\begin{equation}
\kappa=\bigcap_i \kappa_i
\end{equation}
The cells of the collection must be disjoint, because they have the same color.
Of interest here is the case where the $m$ cells $c_i$ have all dimension $e+1$ (as they correspond to gauge generators).
Suppose that $|\kappa|=d'$, so that the cells in the intersection have dimension $d'$.
Since there are a total of $D+1$ colors, $m$ cannot have an arbitrary value, but rather
\begin{equation}
(e+1-d')m\leq (m-1)(D+1-d')
\end{equation}
Indeed, consider the collection of pairs $(i,r)$ where $i=1,\dots, m$ and $r$ is one of the colors in $\kappa_i-\kappa$.
The left hand side is the total number of such pairs.
The right hand side is the maximum number of pairs that we could form with such $r$, which cannot be shared by the $m$ cells $c_i$ and therefore can appear at most $m-1$ times.
Thus the inequality, which can be restated as
\begin{equation}\label{eq:simplify}
d'\geq D-m \bar e+1.
\end{equation}
Going back to \eqref{eq:condition_empty}, the case $m=n$ will be satisfied if and only if the intersection is composed of cells of dimension at least 1.
For this to hold, the right hand side of \eqref{eq:simplify} hast to be greater or equal than one, \emph{i.e} the inequality \eqref{eq:condition_D} must hold.
Naturally, the conditions \eqref{eq:condition_empty} must be satisfied also for $m<n$, but when \eqref{eq:condition_D} holds
\begin{equation}
d'\geq D-m \bar e+1\geq (n-m)\bar e+1\geq n-m+1.
\end{equation}
Thus the conditions \eqref{eq:condition_empty} are satisfied if all $d$-cells have a number of vertices multiple of $2^d$.
This motivates the following definition:

\vspace{.2cm}
\noindent \emph{A $D$-colex is perfect if every cell has  $0 \mod 2^d$ vertices, where $d$ is the dimension of the cell.}
\vspace{.2cm}

In~\cite{bombin:2013:self} the problem of the existence of perfect colexes was left open.
The rest of this appendix is devoted to show that one can always obtain a perfect colex from a generic one by transforming the lattice in the neighborhood of a collection of vertices $T$. 
The next appendix in turn shows that the actual substitution of the generic colex by its perfect counterpart is unnecessary. 
Rather, a transversal $\rot n$ is recovered by inverting the rotation at those qubits in the set $T$.

\subsection{Construction}

\begin{figure}
  \centering
  \includegraphics[width=8cm]{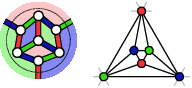}
  \caption{
  Direct and dual pictures of the transformation applied to each element of $T$ for $D=2$.
 In the 2-colex picture (left) the neighborhood of each vertex in $T$ is substituted by a 2-ball (with contour the dotted line) obtained from a colored cube by removing the neighborhood of a vertex.
  In the simplicial picture (right) each 2-simplex in $T$ is divided in 7 pieces.
  This adds 3 new vertices (those in the interior), each of a different color.
  }
\label{fig:divide}
\end{figure}

Crucial to the construction of perfect colexes are the `simplest' colexes described in the previous appendix, which \emph{are} perfect.
Recall that the boundary of a $(D+1)$-hypercube is a $D$-colex, from which the `simplest' punctured $D$-colex is obtained by removing a vertex and all the cells containing it.
Here instead we consider removing a small neighborhood of a vertex, so that we are left with a $D$-dimensional ball that contains $2^{D+1}-1$ vertices and in which all `complete' $d$ cells have $2^d$ vertices, while the `incomplete' ones are lacking the removed vertex and thus keep only $2^d-1$.
The procedure to make a generic colex into a perfect one involves substituting the neighborhood of each of the vertices in a collection $T$ by such $D$-balls, as in Fig.~\ref{fig:divide} (this is actually a connected sum with a neat combinatorial description, see~\cite{bombin:2007:branyons}).
Therefore, a given $d$ cell $c$ with vertex set $V_c$ in the original colex gives rise to a $d$-cell $c'$ with vertex set $V_{c'}$ such that
\begin{equation}\label{eq:changeV}
|V_{c'}|=|V_c|+(2^{d}-2)|T\cap V_c|.
\end{equation}
All the new $d$-cells added to the colex have $2^d$ vertices.

The procedure is generically valid only for either spherical $D$-colexes, which yield trivial codes, or punctured $D$-colexes, which are the ones of interest here.
Notice however that in order for the above to make sense we need to be sure that every vertex has the right kind of neighborhood, which is true in the bulk but not on the boundary for the punctured colex.
Fortunately this is not an important issue. 
Indeed, it suffices to make the colex into an spherical one by adding a 0-cell in the usual way, perform the changes at $T$ vertices, and then remove the added vertex together with all the cells that contain it.

It might be worth describing the dual picture, in which $T$ is a set of $D$-simplices. 
Each simplex in $T$ is divided in $2^{D+1}-1$ pieces:
there are $D+1$ new vertices, and for each color set $\kappa$ with $0<|\kappa|<D+1$ there is a new $D$-simplex that has as vertices (i) the new vertices with labels $\kappa$ and (ii) the vertices of the removed simplex with colors in $\bar\kappa$.
Fig.~\ref{fig:divide} illustrates the dual picture for $D=2$.
Notice that this is nothing but the construction that was already used in the main text to `close' $M$ to form a sphere.

\begin{figure}
  \centering
  \includegraphics[width=8.3cm]{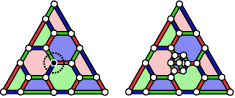}
  \caption{
A punctured 2-colex (left) and a perfect punctured 2-colex (right) obtained from the former by changing the lattice in the neighborhood of the vertex marked in black.
}
\label{fig:perfect}
\end{figure}

It is convenient to first consider the $D=2$ case.
Fig.~\ref{fig:perfect} illustrates the procedure in a 2-colex for which $T$ is composed of a single element, \emph{i.e.} it is enough to substitute the neighborhood of a single vertex.
In the original 2-colex there are 3 faces that have 6 vertices.
In the transformed one, each of these faces has gained 2 vertices, so that they have $8\equiv 0\mod 4$ vertices, as required.

It is apparent what the $T$ set of vertices should satisfy in the $D=2$ case.
`Good' faces, those with $0\mod 4$ vertices, must have an even number $v$ of vertices in $T$, so that they end up with 
\begin{equation}
0+2v\equiv 0 \mod 4
\end{equation}
vertices and stay good.
`Bad' faces, those with $2\mod 4$ vertices, must have an odd number $v$ of vertices in $T$, so that they end up with 
\begin{equation}
2+2v\equiv 0 \mod 4
\end{equation}
vertices and become good.
But, is there always such a set $T$? How do we find it?

At this point it is useful to consider a different problem: error correction.
In the $D=2$ case, there is a single color code: $(1,1)$. 
Its stabilizer generators are attached to faces.
Given an error of the form $X_T$, where $T$ is some set of qubits/vertices, its syndrome amounts to the collection of faces $f$ such that $X_T$ and $Z_f$ anticommute, where $Z_f$ stands for the $Z$ stabilizer at $f$.
These syndrome faces $f$ are simply those that contain an odd number of vertices from the set $T$.
\emph{I.e.} the syndrome of $X_T$ corresponds to those faces that gain $2\mod 4$ vertices when the vertices of $T$ are subject to the above geometrical transformation.
This allows to answer both questions above.
Regarding existence, in the case of punctured $2D$ color codes all syndrome sets are possible~\cite{bombin:2006:2dcc}.
To find $T$ it suffices to solve a set of binary linear equations, just the same way that one can find the possible errors from the error-syndrome.


The stage is set to show how the construction works for general punctured $D$-colexes.
There are two steps to this.
First, it is possible to find a set of vertices $T$ such that the resulting $D$-colex has only good 2-cells as above.
Second, it turns out that the resulting colex is perfect.

The existence of $T$ is not obvious.
The reason for this is that for $D>2$ not every syndrome is possible.
In particular, consider the $(D-1,1)$ conventional color code on the given $D$-colex: it has $Z$ stabilizer generators attached to 2-cells.
These 2-cell operators $Z_f$ are not independent.
Instead, in a punctured code they satisfy only local constraints that have their origin in the structure of $3$-cells~\cite{bombin:2007:branyons}.
Namely, every 3-cell $c$ has 3 colors, and thus is composed of 2-cells with 3 different color sets $\kappa_i$, $i=1,2,3$.
Let $F_\kappa$ be the set of 2-cells of $c$ with color $\kappa$, where $\kappa\in\sset{\kappa_1,\kappa_2,\kappa_3}$.
By definition, every vertex of $c$ belongs exactly to one 2-cell in each $F_\kappa$.
A trivial consequence is
\begin{equation}\label{eq:constraint_local}
\prod_{f\in F_\kappa} Z_f = \prod_{f\in F_{\kappa'}} Z_f,\qquad \kappa,\kappa' \in\sset{\kappa_1,\kappa_2,\kappa_3}.
\end{equation}
In words: the number of 2-cell operators $Z_f$ with $f\in F_\kappa$ and negative syndrome is even if and only if the number of 2-cell operators $Z_f$ with $f\in F_{\kappa'}$ and negative syndrome is even.

Fortunately, the constraints \eqref{eq:constraint_local} are no obstacle for the present purpose because bad 2-cells satisfy them.
Namely, if among the 2-cells in $F_\kappa$ there are $m_\kappa$ bad ones, then $|V_c|$, the total number of 0-cells in $c$, satisfies
\begin{equation}\label{eq:count_c}
|V_c|\equiv 2m_\kappa \equiv 2m_{\kappa'} \mod 4.
\end{equation}
This follows again from the fact that every vertex of $c$ belongs exactly to one 2-cell in each of the sets $F_\kappa$:
the elements of $V_c$ can be counted by adding the number of vertices in each 2-cell of $F_\kappa$ for any given $\kappa$: the result is always the same.
But \eqref{eq:count_c} yields
\begin{equation}
m_\kappa \equiv m_{\kappa'} \mod 2,
\end{equation}
which means that a syndrome such that 2-cells have eigenvalue $+1/-1$ if they are respectively good/bad does exist because the constraints \eqref{eq:constraint_local} are satisfied.
There exists $T$ such that the set of 2-cell operators $Z_f$ that anticommute with $X_T$ is the same as the set of bad 2-cells.

It only remains to show that every $D$-colex with no bad 2-cells is perfect.
This can be done recursively.
Assume that for a given $D$-colex with no bad 2-cells every $d$-cell has $0\mod 2^d$ vertices.
The aim is to show that every $(d+1)$-cells has $0\mod 2^{d+1}$ vertices.

First, every such $(d+1)$-cell has a boundary that is a spherical $d$-colex.
From this spherical $d$-colex one can build a new `shrunk' $d$-dimensional lattice by shrinking all the $d$-cells with a given color set $\kappa$ to a point~\cite{bombin:2007:branyons}:
\begin{itemize}
\item The 0-cells of the shrunk lattice correspond to $d$-cells with color set $\kappa$, or $\kappa$-cells, of the $d$-colex.
\item The 1-cells of the shrunk lattice correspond to 1-cells of color $r$ of the $d$-colex, or $r$-cells, where $r$ is the only color with $r\not\in\kappa$.
\item The 2-cells of the shrunk lattice correspond to 2-cells with color sets $\kappa'$ such that $r\in\kappa'$.
In particular, the shrunk 2-cell has an even number of 1-cells in its boundary because the original 2-cell has $0\mod 4$ 1-cells and half of them are $r$-cells.
\end{itemize}

Since the homology of the sphere is trivial and all the 2-cells have an even number of edges, the graph formed by the 0-cells and 1-cells of the shrunk lattice is bipartite.
Indeed, any closed path must have an even number of vertices and edges because (i) as a $\mathbf Z_2$ 1-chain it has no boundary and thus must be a sum of boundaries of 2-cells and (ii) the sum of 1-chains with an even number of edges must have an even number of edges.

Therefore, the set of $\kappa$-cells in the $d$-colex is the disjoint union of two sets $A$ and $B$, and every $r$-cell in the $d$-colex shares a vertex with exactly one element of the set $A$.
Since every $\kappa$-cell in $A$ has $0\mod 2^d$ vertices, it follows that there are $0\mod 2^d$ $r$-cells and therefore $0\mod 2^{d+1}$ vertices in the $d$-colex (every vertex belongs exactly to one $r$-cell, and each $r$-cell has 2 vertices).

\section{Transversal $\rot n$ in color codes}\label{app:transversal_cc}

According to the discussion in appendix \ref{app:perfect_motivation}, if condition \eqref{eq:condition_D} is satisfied then $\rot n$ can be implemented transversally in a $(d,e)$ gauge color code as long as the colex satisfies the following property for some set $T$ of vertices

\vspace{.2cm}
\noindent\emph{Every $d$-cell $c$ of the $D$-colex has a set $V_c$ of vertices with 
\begin{equation}\label{eq:cellsT}
|V_c|_T \equiv 0\mod 2^{d}.
\end{equation}
}

\noindent Indeed, \eqref{eq:assumption} can be recovered from \eqref{eq:cellsT} using the fact that for $A,B$ disjoint sets
\begin{equation}
|A\sqcup B|_T=|A|_T+|B|_T
\end{equation}

The trick to get \eqref{eq:cellsT} is to choose $T$ as in the previous appendix. 
Indeed, one can compare the given colex with the perfect one obtained by transforming the neighborhoods of the $T$ vertices.
The cell $c$ maps to a new cell $c'$ in the perfect colex and, in the notation of \eqref{eq:changeV},
\begin{equation}
|V_{c}|_T\equiv |V_{c}|-2|T\cap V_c| \equiv |V_{c'}|\equiv 0 \mod 2^d,
\end{equation}
where the first equivalence is by definition of $|T|$, the second follows from \eqref{eq:changeV}, and the third is due to the perfection of the final colex.

\bibliography{refs,comments}

\end{document}